\documentclass[aps,twocolumn,showpacs,superscriptaddress,floatfix]{revtex4}

\usepackage[utf8]{inputenc}
\usepackage{braket}
\usepackage{amsmath,amssymb,graphicx}
\usepackage{qcircuit}
\usepackage{xcolor}
\usepackage{verbatim}
\usepackage[normalem]{ulem}
\usepackage{graphicx}
\usepackage[rightcaption]{sidecap}

\newcommand{\norm}[1]{\left\lVert#1\right\rVert}
\newcommand{\R}{\mathbb{R}}
\def\bra#1{\mathinner{\langle{#1}|}}
\def\ket#1{\mathinner{|{#1}\rangle}}

\usepackage{float}

\begin{document}
\title{Nearest Centroid Classification on a Trapped Ion Quantum Computer}

\author{Sonika Johri}
\affiliation{IonQ Inc, 4505 Campus Dr, College Park, MD 20740}

\author{Shantanu Debnath}
\affiliation{IonQ Inc, 4505 Campus Dr, College Park, MD 20740}

\author{Avinash Mocherla}
\affiliation{QC Ware, Palo Alto, USA and Paris, France}
\affiliation{UCL, Centre for Nanotechnology, London, UK}

\author{Alexandros Singh}
\affiliation{QC Ware, Palo Alto, USA and Paris, France}
\affiliation{Universit\'e Sorbonne Paris Nord, France}

\author{Anupam Prakash}
\affiliation{QC Ware, Palo Alto, USA and Paris, France}

\author{Jungsang Kim}
\affiliation{IonQ Inc, 4505 Campus Dr, College Park, MD 20740}

\author{Iordanis Kerenidis}
\affiliation{QC Ware, Palo Alto, USA and Paris, France}
\affiliation{CNRS, University of Paris, France}


\begin{abstract}
Quantum machine learning has seen considerable theoretical and practical developments in recent years and has become a promising area for finding real world applications of quantum computers. 
In pursuit of this goal, here we combine state-of-the-art algorithms and quantum hardware to provide an
experimental demonstration of a quantum machine learning application with provable guarantees for its performance and efficiency. In particular, we design a quantum Nearest Centroid classifier, using techniques for efficiently loading classical data into quantum states and performing distance estimations, and experimentally demonstrate it on a 11-qubit trapped-ion quantum machine, matching the accuracy of classical nearest centroid classifiers for the MNIST handwritten digits dataset and achieving up to $100\%$ accuracy for 8-dimensional synthetic data. 
\end{abstract}

\maketitle

\section{Introduction}

Quantum technologies promise to revolutionize the future of information and communication, in the form of quantum computing devices able to communicate and process massive amounts of data both efficiently and securely using quantum resources. Tremendous progress is continuously being made both technologically and theoretically in the pursuit of this long-term vision.

A primary goal of current quantum computing research is to find real-world applications of the quantum computers that will become available in the coming years. In order to arrive at these first applications, simultaneous progress on both hardware and algorithms is required.

On the one hand, quantum hardware is making considerable advances. Small quantum computers capable of running representative algorithms were first made available in research laboratories, utilizing both trapped ion~\cite{DebnathNature2016,MonzScience2016} and superconducting qubits~\cite{BarendsNature2014,CorcolesNatCom2015}. Performance comparisons among different quantum computer hardware have been made for running a host of quantum computing tasks~\cite{LinkePNAS2017,MuraliISCA2019}. Access to noisy intermediate-scale quantum (NISQ) computers from several commercial vendors is now available through cloud services. A recent report on achieving quantum supremacy, where a task was performed by a quantum computer that cannot be simulated with any available classical computer~\cite{AruteNature2019}, is an indication that powerful quantum computers will likely be available to researchers in the near future.

At the same time, considerable algorithmic work is underway in order to reduce the resources needed for implementing impactful quantum algorithms and bring them closer to the NISQ era. For example, more NISQ variants of amplitude estimation algorithms, a fundamental quantum procedure that is used in a large number of quantum algorithms, appeared recently \cite{suzuki2020amplitude,tanaka2020amplitude,grinko2019iterative,aaronson2020quantum,qfin2020}. Another example are compilation techniques for optimizing the number of quantum gates that have resulted in a reduction in the number of qubits and gates for the factoring algorithm by many orders of magnitude \cite{gidney2019factor}. 

In this work, we focus on the area of quantum machine learning. There are a number of reasons why machine learning is a good area for trying to find applications of quantum computers. 
First, classical machine learning has proved to be an extremely powerful tool for a plethora of sectors, including Healthcare, Automotive, Manufacturing and Finance, so understanding the enhancements quantum computing can offer to this area will have a major impact.
Second, we already know that fault-tolerant quantum computers with fast quantum access to classical data can provably offer advantages for many different applications such as classification, clustering and recommendation systems \cite{lloyd2014quantum, kerenidis2016quantum, kerenidis2019qmeans, li2019sublinear}. In this work, we will show that there are concrete avenues for reducing the resources needed for implementing some core elements of such algorithms, in particular for loading classical data as quantum states and performing distance estimation between data points. Another interesting point is that it is desirable for classical machine learning algorithms to be robust against noise inherent in the data. To this end, regularization techniques that artificially inject noise into the computation are often used to improve generalization performance for classical neural networks \cite{noiseNN2017}. Thus, one might hope that noisy quantum computers are inherently better suited for machine learning computations than for other types of problems that need precise computations like factoring or search problems. Last, there are many performance measures one may want to improve when it comes to machine learning - in addition to efficiency and accuracy, metrics like interpretability, transparency, and energy consumption are important ones where quantum computing may offer an advantage.

However, there are significant challenges to be overcome to make quantum machine learning practical. First, most quantum machine learning algorithms that offer considerable speedups assume that there are efficient ways to load classical data into quantum states. We address this bottleneck in this paper and describe ways to load a classical data point with logarithmic depth quantum circuits and using a number of qubits equal to the features of the data point. 
Another algorithmic bottleneck is that the solution output by the algorithm is oftentimes a quantum state from which one needs to extract some useful classical information. At times this can be efficient, for example, when an estimate on the distance between two such states or the identification of heavy-hitters is desired. In general, an exponential amount of time may be needed to extract the classical description of the quantum state by performing tomography. 

One also needs to be careful with the efficiency of many of the quantum subroutines used in quantum machine learning, in particular linear algebra subroutines, since their running time depends on a large number of instance-specific parameters that need to be taken into account before claiming any speedups. In particular, most of these speedups will be polynomial and not exponential, and this is also corroborated by quantum inspired algorithms \cite{tang2019quantum}. In any case, we believe the right way to describe these quantum speedups may not be to merely state them as exponential or polynomial, but to quantify the extent of speedup for each application and to see how these theoretical speedups translate in practice.   

Another promising avenue for quantum machine learning pertains to the use of parametrized quantum circuits as analogues of neural networks for supervised learning, in particular for classification \cite{QNN2018, CNN2019}. In fact, classical neural networks achieve extremely high performance for specific classification tasks like image classification, and one hopes that quantum analogues can achieve speedups and further enhance the accuracy of such techniques. Again, one needs to be careful, since we neither have much theoretical evidence that such quantum architectures will be easily trained, nor can we perform large enough simulations to get any convincing practical evidence of their performance (since we do not have large enough quantum hardware and classical simulations incur an exponential overhead). 
For example, architectures that use only a constant depth and only gates between consecutive qubits, while being suitable for near-term quantum computers, cannot really act as a fully connected neural network, since each input qubit can only affect a constant number of output qubits. 
It is also becoming clear that the time to train such quantum variational circuits can be quite large, both because of phenomena such as barren plateaus and also since designing the architectures, choosing cost-functions and initializing the parameters is far more complex and subtle than one may naively think \cite{QCNNplateaus2020, VQCA2020, Train2020}. Further work is needed to understand the power and limitations of variational quantum circuits for machine learning applications.

Our work is a collaboration between quantum hardware and software teams that advances the state-of-the-art of quantum machine learning implementations, bringing potential applications closer to reality. Even though the scale of the implementation remains a proof of concept, our work makes significant progress towards unblocking a number of theoretical and practical bottlenecks. In particular, we look at classification, one of the canonical problems in supervised learning with a vast number of applications. 
In classification, one uses a labelled dataset (for example, emails labelled as Spam or Not Spam) to fit a model which is then used to predict the labels for new data points (for example, predict whether a new email should be labelled Spam or Not Spam).
There are many different ways to perform classification that one can broadly place in two main categories. 

The first way is similarity-based learning, where a notion of similarity between data points is defined (e.g. the Euclidean distance between data points seen as vectors) and points are classified together if they are similar. Well-known similarity-based algorithms are the Nearest Centroid, $k$-Nearest Neighbors, Support Vector Machines, etc. 
The second way is based on deep learning techniques, in particular on different types of neural networks (fully connected, convolutional, recurrent, etc.). Here, the corpus of labelled data is used in order to train the weights of a neural network so that once trained it can infer the label of new data. 
Often, especially in cases where there is a large amount of data, neural networks can achieve better performance than more traditional similarity-based methods. On the other hand, similarity-based methods can offer other advantages, including provable performance guarantees and also properties like interpretability and transparency, which are becoming increasingly important in many sectors with sensitive data and decision making. 

Here, we focus on demonstrating a quantum analogue of the Nearest Centroid algorithm, a simple similarity-based classification technique. The Nearest Centroid algorithm is a good baseline classifier that offers interpretable results, nevertheless, its performance deteriorates when the data points are far away from belonging to convex classes with similar variances. 
The algorithm takes as input a number of labelled data points, where each data point belongs to a specific class. The model fitting part of the algorithm is very simple and it involves computing the centroids, i.e. the barycenters of each of the sets. 
Once the centroids of each class are found, then a new data point is classified by finding the centroid which is nearest to it in Euclidean distance and assigning the corresponding label.

In our work, we design a quantum Nearest Centroid algorithm, by constructing quantum procedures for loading the classical data as quantum states and performing a distance estimation procedure.
We demonstrate the quantum Nearest Centroid algorithm on up to 8 qubits of a trapped ion quantum processor and achieve accuracies comparable to corresponding classical classifiers on real datasets, as well as 100\% accuracies on synthetic data. To our knowledge, this is the largest and most accurate classification demonstration on quantum computers. Importantly, we develop an error mitigation technique and noise model analysis that prove this performance will continue to hold as the problem size scales up.

{\em Related experimental work.}
We describe here some previous work on classification experiments on quantum computers, in particular with neural networks. 
In fact, there is a fast growing literature on variational
methods for classification on small quantum computers \cite{QNN2018,CNN2019,Image2020,Semisupervised2020,Polyadic2020,Dressed2020,Supervised2018,Hierarchical2018} of which we briefly describe those that also include hardware implementations. 
In \cite{Supervised2018}, the authors provide  binary classification methods based on variational quantum circuits. The classical data is mapped into quantum states through a fixed unitary transformation and the classifier is a short variational quantum circuit that is learned through  stochastic gradient descent. A number of results on synthetic data are presented showing the relation of the method to Support Vector Machines classification and promising performance for such small input sizes.  
In \cite{Hierarchical2018} the authors provide a number of different classification methods, based on encoding the classical data in separable qubits and performing different quantum circuits as classifiers, inspired by Tree Tensor Network (TTN) and Multi-Scale Entanglement Renormalization Ansatz (MERA) circuits. A 4-qubit hardware experiment for a binary classification task between two of the IRIS dataset classes was performed on an IBM machine with high accuracy. 
In \cite{Polyadic2020}, the authors performed 2-qubit experiments on IBM machines and with the IRIS dataset. They reported high accuracy for a subset of the dataset after training a quantum variational circuit for more than an hour and 3 million circuit runs. \\

The remainder of the paper is organized as follow: Section II explains the algorithm and software development. The experimental results and noise model are described in Section III. We end with a discussion in Section IV.

\section{Algorithm and Software}

In this section, we describe the algorithm and software tools we used to implement quantum classification. 

\subsection{Data loaders and distance estimation}

We start by describing our data loaders \cite{Ker2020}. Being able to load classical data as quantum states that can be efficiently used for further computation is an important step for machine learning applications, since on the one hand, data is, and will likely remain, predominantly classical, and on the other, 
most quantum applications are based on efficient quantum access to classical data, whether these are linear system solvers, convex optimization, unstructured search, etc.

Let us start by defining more precisely what we mean by a data loader. A data loader is a procedure that, given access to a classical data point $x = (x_1,x_2,\ldots, x_d) \in \R^d$, pre-processes the classical data efficiently, i.e. reading the data once and spending $\widetilde{O}(d)$ time overall, and outputs a parametrized quantum circuit of size $O(d)$ but of depth only $O(\log d)$, 
that prepares quantum states of the form
\[ 
\frac{1}{\norm{x}}\sum_{i=1}^{d} x_i \ket{i}. 
\]
Here, $\ket{i}$ is some representation of the numbers $1$ through $d$ (we will use a unary representation in the experiment but we describe other representations as well).

Let us remark that to calculate the efficiency of our algorithms, we assume that quantum computers will have the ability to perform gates on different qubits in parallel. This is possible to achieve in most technologies for quantum hardware, including ion traps \cite{parallelgates}. The specific quantum states we consider, also called ``amplitude encodings'' are not the only possible way to load classical data into quantum states but they are by far the most interesting in terms of the quantum algorithms that can be applied to them. For example, they are the states one needs in order to start the quantum linear system solver procedure. Note also, that if we want to exactly load classical data points with $d$ dimensions then we have $d-1$ degrees of freedom for defining such quantum states (since they are normalized to be unit vectors), so we need a circuit of size at least $d-1$. In fact our circuits have exactly $d-1$ two-qubit parametrized gates as we will see below. One also needs to keep track of the norm of the vectors which can be easily computed during the pre-processing.

There have been several proposals for acquiring fast quantum access to classical data that loosely go under the name of QRAM (Quantum Random Access Memory). A QRAM, as described in  \cite{QRAM2007,QRAM2008}, in some sense would be a specific hardware device that could ``natively'' access classical data in superposition, thus having the ability to create quantum states like the one defined above in logarithmic time. Given the fact that such  specialized hardware devices do not yet exist, nor do they seem to be easy to implement, there have been proposals for using quantum circuits to perform similar operations. For example, a circuit to perform the bucket brigade architecture was defined in \cite{bucket2015}, where a circuit with $O(d)$ qubits and $O(d)$ depth was described and also proven to be robust up to a level of noise. A more ``brute force'' way of loading a $d$-dimensional classical data point is through a multiplexer-type circuit, where one can use only $O(\log d)$ qubits but for each data point one needs to sequentially apply $d$ $\log d$-qubit-controlled gates, which makes it quite impractical. Another direction is loading classical data using a unary encoding. This was used in \cite{unary2019} to describe finance applications, where the circuit used $O(d)$ qubits and had $O(d)$ depth. A parallel circuit for specifically creating the $W$ state also appeared in \cite{Wparallel2019}.

The loader we will use for our implementation is a ``parallel'' unary loader that loads a data point with $d$ features, each of which can be a real number, with exactly $d$ qubits, $d-1$ parametrized 2-qubit gates, and depth $\log d$. The parallel loader can be viewed as a part of a more extensive family of loaders with $Q$ qubits and depth $D$, with $QD=O(d\log d)$, in particular one can define an optimized loader with $2\sqrt{d}$ qubits and $\sqrt{d}\log d$ depth with $(d-1)$ two- and three-qubit gates in total \cite{Ker2020}. 

Note that the number of qubits we use, one per feature, is the same as in most quantum variational circuit proposals (e.g. \cite{QNN2018,Hierarchical2018}).
One last remark before we give our construction is that here we are talking about loading the exact classical data into quantum states, which is necessary for tasks like classifying specific data points. For other tasks, like training neural networks, one could potentially use classical or quantum techniques to generate ``similar'' data instead of loading the exact data.  One can also perform classical pre-processing of the ``raw'' data, for example dimensionality reduction techniques, before creating the data that one needs to load into quantum states, which is compatible with our techniques and is used for some of the experiments. 

\paragraph{Data loader construction}

We start by a procedure that given access to a classical data point $x = (x_1,x_2,\ldots, x_d) \in \R^d$, pre-processes the classical data efficiently, i.e. spending only $\widetilde{O}(d)$ total time, in order to create a set of parameters $\theta = (\theta_1,\theta_2,\ldots, \theta_{d-1}) \in \R^{d-1}$, that will be the parameters of the $(d-1)$ two-qubit gates we will use in our quantum circuit. In the pre-processing, we also keep track of the norms of the vectors.  

It is important here to notice that the classical memory is accessed once (we use read-once access to $x$) and the parameters $\theta$ are ``stored inside the quantum circuit" (as the parameters of the quantum gates), which means that if we need to perform many operations with the specific data point (which is the case here and, for example, in training neural networks), we do not need to access the classical memory again, we just need to re-run the quantum circuit that already has the parameters in place. 

Let us now describe how to find and store these parameters $\theta$. 
The data structure for storing $\theta$ is in fact the one used in \cite{kerenidis2016quantum}, but note that there we assumed that we have quantum access to these parameters (in the sense of being able to query these parameters in superposition) while here we will compute and store these parameters classically and also encode them as the parameters of the gates used in the quantum circuit. 

At a high level, we think of the coordinates $x_i$ as the leaves of a binary tree of depth $\log d$. The parameters $\theta$ correspond to the values of the internal tree nodes, starting form the root and going towards the leaves. 

We first consider the parameter series $(r_1,r_2, \ldots,r_{d-1})$. For the last $d/2$ values $(r_{d/2}, \ldots, r_{d-1})$, we define an index $j$ that takes values in the interval $[1, d/2]$ and define the values as
\[
r_{d/2+j-1} = \sqrt{x_{2j}^2+x_{2j-1}^2}
\]

For the first $d/2-1$ values, namely the values of  $(r_1,r_2,\ldots,r_{d/2-1})$, and for $j$ in $[1,d/2]$, we define

\[
r_{j} = \sqrt{r_{2j+1}^2+r_{2j}^2}
\]

We can now define the set of angles $\theta = (\theta_1,\theta_2,\ldots, \theta_{d-1})$ in the following way. We start by defining the last $d/2$ values $(\theta_{d/2}, \ldots, \theta_{d-1})$. To do so, we define an index $j$ that takes values in the interval $[1, d/2]$ and define the values as
\begin{eqnarray*}
\theta_{d/2+j-1} & = & \arccos\bigg(\frac{x_{2j-1}}{r_{d/2+j-1}}\bigg) , \mbox{ if } x_{2j} \mbox{ is positive }\\
\theta_{d/2+j-1} & = & 2\pi - \arccos\bigg(\frac{x_{2j-1}}{r_{d/2+j-1}}\bigg) , \mbox{ if } x_{2j} \mbox{ is negative. }
\end{eqnarray*}

For the first $d/2-1$ values, namely the values for $j \in [1,d/2]$, we define
\[
\theta_{j} = \arccos\bigg(\frac{r_{2j}}{r_{j}}\bigg) 
\]

Note that we can easily perform these calculations in a read-once way, where for every $x_i$ we update the values that are on the path from the $i$-th leaf to the root. This also implies that when one coordinate of the data is updated, then the time to update the $\theta$ parameters is only logarithmic, since only $\log d$ values of $r$ and of $\theta$ need to be updated. 

Now that we have found the parameters that we need for our parametrized quantum circuit, we can define the architecture of our quantum circuit. It will use $d$ qubits, $d-1$ two-qubit gates, $\log d$ depth, and will resemble a binary tree architecture. For convenience, we assume that $d$ is a power of 2. 

We will use a gate that has appeared with small variants with different names as partial SWAP, or fSIM, or Reconfigurable BeamSplitter, etc.  We call this two-qubit parametrized gate $RBS(\theta)$ and we define it as 
\begin{equation} \label{RBS}
RBS(\theta) = \left( \begin{array}{cccc}
1 & 0 & 0 & 0 \\
0 & \cos \theta & \sin \theta & 0 \\
0 & -\sin\theta & \cos\theta & 0 \\
0 & 0 & 0 & 1  \end{array} \right)
\end{equation}

One can define the above gate with imaginary off-diagonal elements and in fact we will use that definition when we implement this on the hardware but we keep this definition here for ease of exposition. We can think of this gate as a simple rotation by an angle $\theta$ on the two-dimensional subspace spanned by the vectors $\{\ket{10},\ket{01}\}$ and an identity in the other subspace spanned by $\{\ket{00},\ket{11}\}$. A different way would be to think of a single photon entering one of the two input modes of a reconfigurable beam-splitter and getting split into the two output spatial modes with a ratio depending on the parameter $\theta$. We denote by $RBS^{\dagger}(\theta)$ the adjoint gate for which we have $RBS^{\dagger}(\theta)=RBS(-\theta)$.

\begin{SCfigure}[0.9][h]
\label{loader}
\caption{The data loader circuit for an 8-dimensional data point. The angles of the $RBS(\theta)$ gates starting from left to right and top to bottom correspond to $ (\theta_1,\theta_2,\ldots, \theta_{7})$.}
\includegraphics[width=0.2\textwidth]{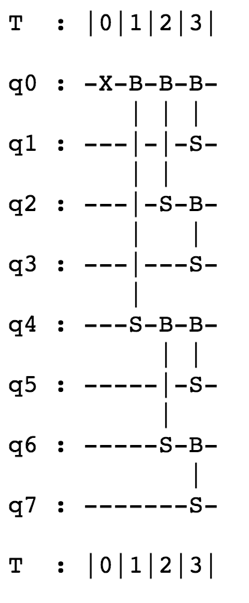}
\end{SCfigure}

We can now describe the circuit itself. We start by putting the first qubit in state $\ket{1}$, while the remaining $d-1$ qubits remain in state $\ket{0}$. Then, we use the first parameter $\theta_1$ in an $RBS$ gate in order to ``split'' this `1' between the first and the $d/2$-th qubit. Then, we use the next two parameters $\theta_2,\theta_3$ for the next layer of two $RBS$ gates, where again in superposition we ``split'' the `1' into the four qubits with indices $(1,d/4,d/2,3d/4)$ and this continues for exactly $\log d$ layers until at the end of the circuit we have created exactly the state
\begin{equation}\label{state}
\ket{x} = \frac{1}{\norm{x}}\sum_{i=1}^{d} x_i \ket{e_i} 
\end{equation}
where the states $\ket{e_i}$ are a unary representations of the numbers $1$ to $d$, using $d$ qubits. The circuit appears in Fig. \ref{loader}.

An interesting extension of our loader is that we can trade off qubits with depth and keep the number of overall gates $d-1$ (in this case, both $RBS$ and controlled-$RBS$ gates). For example, we can use $2\sqrt{d}$ qubits and $O(\sqrt{d}\log d)$ depth \cite{Ker2020}. The circuit is quite simple, if one thinks of the $d$ dimensional vector as a $\sqrt{d}\times \sqrt{d}$ matrix. Then, we can index the coordinates of the vector using two registers (one each for the row and column) and create the state

\[ 
\ket{x} = \frac{1}{\norm{x}}\sum_{i,j=1}^{\sqrt{d}} x_{ij} \ket{e_i}\ket{e_j} 
\]

\begin{figure}[h]
    \centering
    \includegraphics[width=0.5\textwidth]{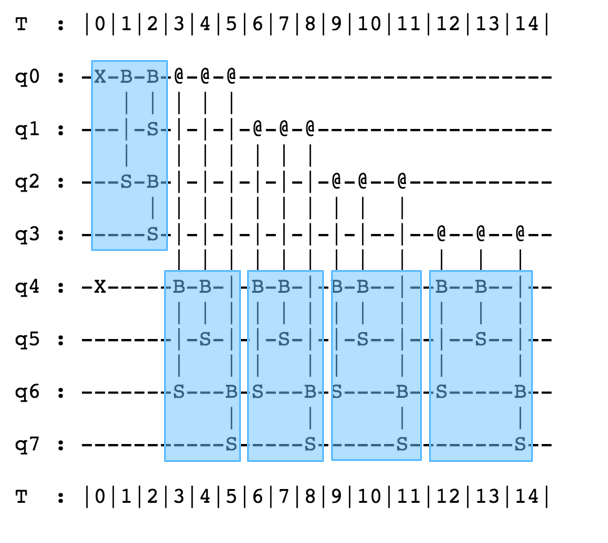}
    \caption{An optimized loader for a 16-dimensional data point, seen as a $4\times4$ matrix. The blue boxes correspond to the parallel loader from Fig. \ref{loader} and its controlled versions. }
    \label{opt_loader}
\end{figure}

For this circuit, we find, in the same way as for the parallel loader, the values $\theta$ and then create the following circuit in Figure \ref{opt_loader}. We start with a parallel loader for a $\sqrt{d}$-dimensional vector using the first $\sqrt{d}$ angles $\theta$ (which corresponds to a vector of the norms of the rows of the matrix) and then, controlled on each of the  $\sqrt{d}$ qubits we perform a controlled parallel loader corresponding to each row of the matrix. 

Notice that naively the depth of the circuit is $O(d\log d)$, but it is easy to see that one can interleave the gates of the controlled-parallel loaders to an overall depth of $O(\sqrt{d}\log d)$. We will not use this circuit here but such circuits can be useful both for loading vectors and in particular matrices for linear algebraic computations.

Let us make some remarks about the data loader circuits.
First, we will see in the following sections that the circuits are quite robust to noise and amenable to efficient error mitigation techniques.
Second, we use a single type of two-qubit gate that is native or quasi-native to different hardware platforms. 
Third, we note that the connectivity of the circuit is quite local, where most of the qubits interact with very few qubits (for example $7/8$ of the qubits need at most 4 neighboring connections) while the maximum number of interacting neighbors for any qubit is $\log d$. Here, we take advantage of the full connectivity of the ionQ hardware platform, so we can apply all gates directly. On a grid architecture one would need to embed the circuit on the grid which asymptotically requires no more than doubling the number of qubits.

\paragraph{Distance estimation circuit}

Given two vectors ${x}$ and ${y}$ corresponding to classical data points, the Euclidean distance, namely $l_{xy}=\norm{{x}-{y}}$ \cite{Ker2020} is given by
\begin{align}
    l_{xy}=\sqrt{\norm{{x}}^2+\norm{{y}}^2 -2\norm{{x}}\norm{{y}}c_{xy}},
\end{align}
where $c_{xy}=\langle x|y\rangle$ is the inner product of the two normalised vectors. Here we describe a circuit to estimate $c_{xy}$ which is combined with the classically calculated vector norms to obtain $l_{xy}$.

The power of the data loaders comes from the operations that one can do once the data is loaded into such ``amplitude encoding'' quantum states. In this work, we show how to use the data loader circuits to perform a fundamental operation at the core of supervised and unsupervised similarity-based learning, which is the estimation of the distance between data points. 

In fact, here we will only discuss one of the variants of the distance estimation circuits which works for the case where the inner product between the data points is positive, which is usually the case for image classification where the data points have all non negative coordinates. It is not hard to extend the circuit with one extra qubit to deal with the case of also non-positive inner products. 

The distance estimation circuit for two data points of dimension $d$ uses $d$ qubits, $2(d-1)$ two-qubit parametrized gates, $2\log d$ depth, and will allow us to measure at the end of the circuit a qubit whose probability of giving the outcome $\ket{1}$ is exactly the square of the inner product between the two normalized data points. From this, one can easily estimate the inner product and the distance between the original data points. When the hardware allows deeper quantum operations, then an amplitude estimation procedure can be used to decrease the number of samples one needs to perform this estimation. For our experiments, we directly repeatedly measured the output state between 500-1000 times in order to get an estimate of the inner product. 

\begin{SCfigure}[0.9][h]
\label{distance}
\caption{The distance estimation circuit for two 8-dimensional data points. The circuit in time steps $0-3$ corresponds to the parallel loader for the first data point and the circuit in time steps $4-6$ corresponds to the inverse parallel loader circuit for the second data point (excluding the last $X$ gate).}
\includegraphics[width=0.25\textwidth]{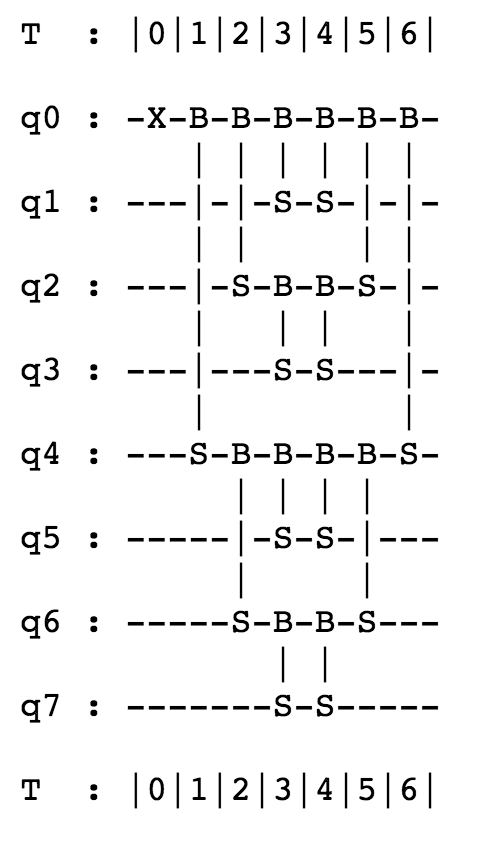}
\end{SCfigure}

The distance estimation circuit is shown in Fig. \ref{distance}, and it consists of two parts, the first is the data loader circuit for the first data point, and the second part is the adjoint data loader circuit for the second data point (without the $X$ gate), where we recall that for the adjoint $RBS^{\dagger}(\theta)=RBS(-\theta)$. We can easily see that the probability the first qubit is measured in state $\ket{1}$ is exactly the square of the inner product of the two data points. 

After the first part, the state of the circuit is $\ket{x}$, as in Eq. \ref{state}. One can rewrite this state in the basis $\{ \ket{y},\ket{y^\bot} \}$ as
\[
\braket{x,y}\ket{y} + \sqrt{1 - |\braket{x,y}|^2 } \ket{y^\bot}
\]
Once the state goes through the inverse loader circuit for $y$ the first part of the superposition gets transformed into the state $ \ket{e_1}$ (which would go to the state $\ket{0}$ after an $X$ gate on the first qubit), and the second part of the superposition goes to a superposition of states $\ket{e_j}$ orthogonal to $\ket{e_1}$ 
\[
\braket{x,y}\ket{e_1} + \sqrt{1 - |\braket{x,y}|^2 } \ket{e_1^\bot}
\]

It is easy to see that after measuring the circuit (either all qubits or just the first qubit), the probability of getting $\ket{1}$ in the first qubit is exactly the square of the inner product of the two data points. 

We can also notice a simplification we can make in the circuit that will reduce the number of gates and depth. In the middle of the circuit, there are pairs of $RBS$ gates that are applied to the same consecutive qubits. Each such pair of two gates can be combined to one gate whose parameter $\theta$ is just equal to $\theta_1+\theta_2$, where $\theta_1$ is the parameter of the first gate and $\theta_2$ is the parameter of the second gate. 

\begin{SCfigure}[0.9][h]
\label{distance_opt}
\caption{The optimized distance estimation circuit for two 8-dimensional data points. The middle layer at time step 3 corresponds to the two merged middle layers from Fig. \ref{distance} at time steps 3 and 4.}
\includegraphics[width=0.25\textwidth]{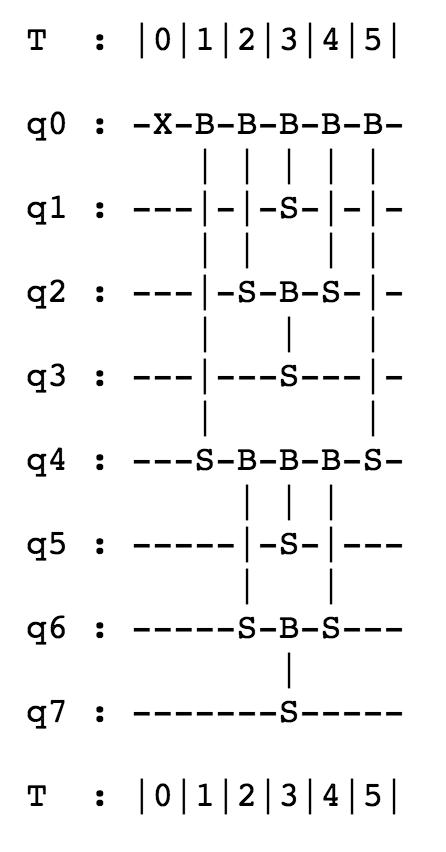}
\end{SCfigure}

This reduces the number of gates of the circuit to $3d/2 - 2$ and reduces the depth by one. The final circuit used in our application is in Fig. \ref{distance_opt}.

\subsection{The Quantum Nearest Centroid classifier}

\subsubsection{Algorithm and software development}

We now have all the necessary ingredients to implement the quantum Nearest Centroid classification circuit. As we have said, this is but a first, simple application of the above tools which can readily be used for other machine learning applications such as nearest neighbor classifiers or k-means clustering. Let us start by briefly defining the Nearest Centroid algorithm in the classical setting. The first part of the algorithm is to use the training data to fit the model. This is a very simple operation of finding the average point of each class of data, meaning one adds all points with the same label and finds the ``centroid'' of each class. This part will be done classically and one can think of this cost as a one-time offline cost. 

In the quantum case, one will still find the centroids classically and then also pre-process them to find the parameters for the gates of the data loader circuits for each one of them and their norms. This does not change the asymptotic time of this step. 

We will now look at the second part of the Nearest Centroid algorithm which is the ``predict'' phase. Here, we want to assign a label to a number of test data points and for that we first estimate the distance between each data point and each centroid and for each data point we assign the label of the centroid which is nearest to it. 

The quantum Nearest Centroid is rather straightforward, it follows the steps of the classical algorithm apart from the fact that whenever one needs to estimate the distance between a data point and a centroid, we do this using the distance estimator circuit defined above. 

The development of the quantum software followed one of the most popular classical ML libraries, called scikit-learn (https://scikit-learn.org/), where a classical version of the Nearest Centroid algorithm is available. In the code snippet below we can see how one can call the quantum and classical Nearest Centroid algorithm with synthetic data (one could also use user-defined data) through QCWare's platform Forge, in a jupyter notebook.  

\begin{figure}[h]
    \centering
    \includegraphics[width=\columnwidth]{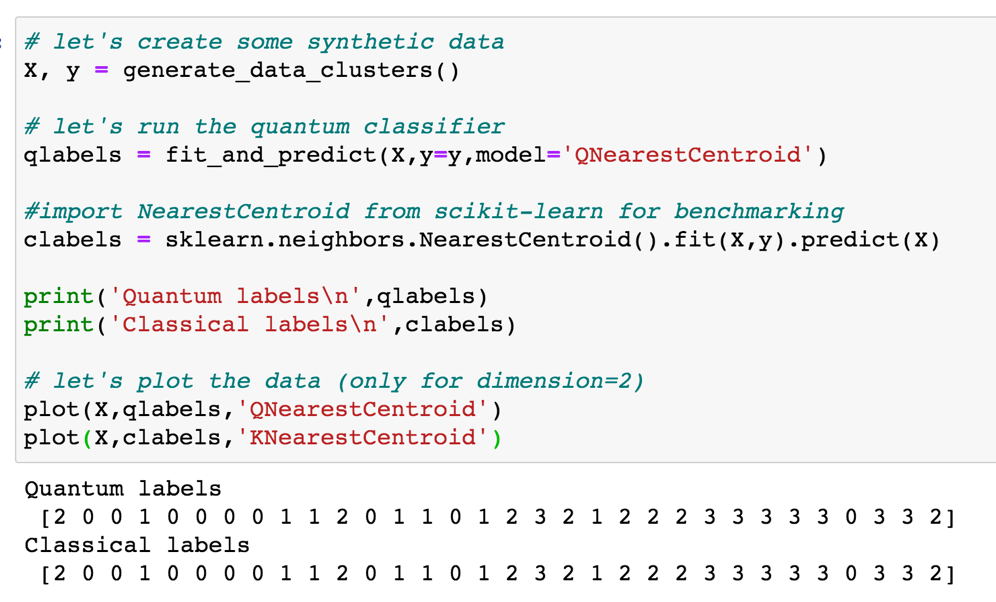}
    \caption{An example of a jupyter notebook that runs the Quantum Nearest Centroid algorithm, as well as the scikit-learn Nearest Centroid algorithm, and prints and plots the results.}
    \label{code}
\end{figure}

The function {\bf fit-and-predict} first classically fits the model. For the prediction it calls a function {\bf distance-estimation} for each centroid and each data point. The {\bf distance-estimation} function runs the procedure we described above, using the function {\bf loader} for each input and returns an estimate of the Euclidean distance between each centroid and each data point. The label of each data point is assigned as the label of the nearest centroid. 

Note that one could imagine more quantum ways to perform the classification, where, for example, instead of estimating the distance for each centroid separately, this operation could happen in superposition. This would make the quantum algorithm faster but also increase the number of qubits needed. We remark also that the distance or inner product estimation procedure can find many more applications such as matrix-vector multiplications during clustering  or training neural networks. 

\subsubsection{Runtime and Scalability}

One of the main advantages of using the Nearest Centroid as a basic benchmark for quantum machine learning is that we fully understand what the quantum algorithm does and how its runtime scales with the dimension of the data and the size of the data set. 

The quantum advantage comes from the distance estimation procedure which is based on the data loader circuits. As we have described, the circuit for estimating the distance of two $d$-dimensional data points has depth $2\log d$. Theoretically, for an estimation of the distance up to $\epsilon$ one needs to run the circuit $N_s=O(1/\epsilon^2)$ times. Note also, that in the future one will be able to use amplitude estimation on top of this circuit in order to reduce the overall time to $O(\log d / \epsilon)$. The accuracy required depends on how well-classifiable our data set is, meaning whether most points are close to a centroid or they are mostly distributed equidistantly from the centroids. This number does not really depend on the dimension of the data set and in all data sets we considered an approximation to the distance up to $0.1$ for most points and $0.03$ for a few difficult to classify points suffices, even for the full-scale MNIST dataset of 784 dimensions. Thus we expect that the number of shots will not significantly change as the problem sizes scale up. 


Since the cost of calculating the circuit parameters is a one-off cost, if we want to estimate the distance between $k$ centroids and $n$ data points all of dimension $d$, then the quantum circuits will need $d$ qubits and the running time would be of the form $O(kd+nd+kn\log (d/\epsilon))$. The first term corresponds to pre-processing the centroids, the second term to pre-processing the new data points and the third term to estimating the distances between each data point and each centroid and assigning a label. The basic classical Nearest Centroid algorithm takes time $O(nkd)$. One can design different classical Nearest Centroid algorithms that also sample using special data structures which will still be quadratically worse than a fully quantum one, at least with respect to the error. 

We note that we do not claim here that the quantum Nearest Centroid procedure is faster than the classical one right now or that it will be in the very near future. Our goal is to measure the performance of the quantum routines outlined above on real hardware and real data. Procedures like the data loader circuits will be useful in the future as an input to much more classically complicated procedures including training neural networks or performing Principal Component or Linear Discriminant Analysis. An important outcome from this is the development of noise models and error mitigation techniques that will also help in improving and predicting the accuracy of the more complicated algorithms.

\section{Experiment}
\subsection{Trapped ion quantum computer}

\begin{figure}
\centering
\includegraphics[width=\columnwidth]{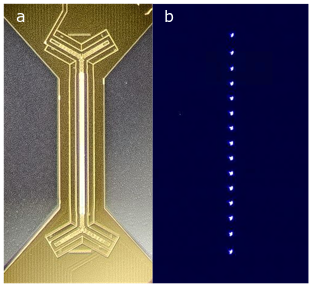}
\caption{The quantum processor consists of a micro-fabricated surface electrode ion-trap (a), which is used to trap a linear chain of Ytterbium ions. (b) A chain of 15 ions is imaged by collecting fluorescence using an optical microscope, where each ion can represent a physical qubit.}
\label{fig:iontrap}
\end{figure}

Our experimental demonstration is performed on an 11-qubit trapped ion processor based on $^{171}$Yb$^{+}$ ion qubits. The device is commercially available through IonQ's cloud service. 

The 11-qubit device is operated with automated loading of a linear chain of ions (see figure \ref{fig:iontrap}), which is then optically initialized with high fidelity. Computations are performed using a mode-locked $355$nm laser, which drives native single-qubit-gate (SQG) and two-qubit-gate (TQG) operations. The native TQG used is a maximally entangling Molmer Sorensen gate.

In order to maintain consistent gate performance, calibrations of the trapped ion processor are automated. Additionally, phase calibrations are performed for SQG and TQG sets, as required for implementing computations in queue and to ensure consistency of the gate perfomance.

\subsubsection{Implementation of the circuits on the ionQ processor}

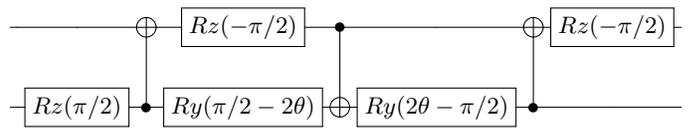
\begin{figure}
\Qcircuit @C=0.3em @R=1.7em {
&\qw &\qw &\targ &\gate{Rz(-\pi/2)} &\ctrl{1}  &\qw &\targ &\gate{Rz(-\pi/2)} &\qw\\
&\qw &\gate{Rz(\pi/2)} &\ctrl{-1} &\gate{Ry(\pi/2-2\theta)} &\targ &\gate{Ry(2\theta-\pi/2)} &\ctrl{-1} &\qw &\qw \\
}
\caption{Circuit to implement $iRBS(\theta)$ gate}
\label{fig:RBS}
\end{figure}

The circuits we described above are built with the $RBS(\theta)$ gates (Eq. \ref{RBS}). To map these gates optimally onto the hardware we will instead use the modified gate
\begin{equation} \label{iRBS}
iRBS(\theta) = \left( \begin{array}{cccc}
1 & 0 & 0 & 0 \\
0 & \cos \theta & -i\sin \theta & 0 \\
0 & -i\sin\theta & \cos\theta & 0 \\
0 & 0 & 0 & 1  \end{array} \right)
\end{equation} 
It is easy to see that the distance estimation circuit stays unchanged. Using the fact that $iRBS(\theta)=\exp(i\theta(\sigma_x\otimes\sigma_x + \sigma_y\otimes \sigma_y))$, we can decompose the gate into the circuit shown in Fig. \ref{fig:RBS} \cite{Vatan2004}. Each CNOT gate can be implemented with a maximally entangling Molmer Sorensen gate and single qubit rotations native to IonQ hardware. 

We run 4 and 8 qubit versions of the algorithm. The 4 qubit circuits have 12 TQG and the 8 qubit circuits have 30 TQG.

\begin{table*}[]
\begin{tabular}{| c | c |}
\hline
 Sampling of points & 0     0     0     0     0     0     0     0     0     0     1     1     1     1     1     1     1     1     1   1     2     2     2     2     2     2     2     2     2     2     3     3     3     3     3     3     3     3  3     3 \\
\hline
Classical NC & 0     0     2     0     0     0     0     0     1     0     1     1     1     1     1     1     1     1     3   1     2     2     2     2     2     2     3     2     2     2     3     3     3     3     3     3     3     3   3     3
\\
\hline
QNC (no mitigation) & 0     0     2     0     0     0     0     0     1     0     1     1     1     1     1     1     1     1     3    1     2     2     2     2     2     2     3     2     2     {\color{red}1}     3     3     3     3     3     3     3     3   3     3  \\
\hline
QNC (with mitigation) &   0     0     2     0     0     0     0     0     1     0     1     1     1     1     1     1     1     1     3   1     2     2     2     2     2     2     3     2     2     {\color{red} 1}     3     3     3     3     3     3     3     3   3     3\\
\hline
\end{tabular}
\caption{Comparison of labels assigned by the different classification schemes for synthetic data with $N_q=4$, $N_c=4$ and $N_s=500$. The labels in red show where the quantum classification differs from the classical one.}
\label{table_data4}
\end{table*}

\begin{table*}[]
\begin{tabular}{| c | c |}
\hline
 Sampling of points & 0     0     0     0     0     0     0     0     0     0     1     1     1     1     1     1     1     1     1     1     2     2     2  2     2     2     2     2     2     2     3     3     3     3     3     3     3     3     3     3 \\
\hline
Classical NC &  0     0     0     0     0     0     0     0     0     0     1     1     1     1     1     1     1     1     1     1     2     2     2  2     2     2     2     2     2     2     3     3     3     3     3     3     3     3     3     3
\\
\hline
QNC (no mitigation) & 0     0     0     0     0     0     0     0     0     0     {\color{red} 0     3     0     0     0     0     3     0     0     0}     2     2     2     2     2     2     2     2     2     2     3     3     3     {\color{red} 2}     3     3    3     3     {\color{red}2}     3\\
\hline
QNC (with mitigation) &  0     0     0     0     0     0     0     0     0     0     1     1     1     1     {\color{red}0}     1     1     1   {\color{red}0}     1     2     2     2     2     2     2     2     2     2     2     3     3     3     {\color{red}2}     3     3   3     3     {\color{red}2}     3\\
\hline
\end{tabular}
\caption{Comparison of labels assigned by the different classification schemes for synthetic data with $N_q=8$, $N_c=4$ and $N_s=1000$. The labels in red show where the quantum classification differs from the classical one.}
\vspace{20pt}
\label{table_data8}
\end{table*}

\subsection{Experimental Results}

We tested the algorithm with synthetic and real datasets. On an ideal quantum computer,  the state right before measurement has non-zero amplitudes only for states within the unary basis,  $\ket{e_i}=\ket{2^i}$, i.e. states in which only a single bit is 1. The outcome of the experiment that is of interest to us is the state $\ket{10..00}$. To estimate the probability of this outcome, the circuit is performed a number of times $N_{s}$ and the probability is calculated as the ratio of the number of times the outcome is the state $\ket{10..00}$ over $N_{s}$. 
On a real quantum computer, each circuit is run with a fixed number of shots to recreate the density matrix probability distribution in the output as closely as possible. Typically, there are many computational basis states other than the ones in the ideal probability distribution that are measured. 
Therefore, we have adopted two different techniques for estimating the desired probability.

\begin{enumerate}
\item No error-mitigation: Measure the first qubit and compute the probability as the ratio of the number of times the outcome is 1 over the total number of runs of the circuit $N_{s}$.
\item Error-mitigation: Measure all qubits and discard all runs of the circuit with results that are not of the form $\ket{2^i}$. Compute the probability as the ratio of the number of times the outcome is $\ket{10..0}$ over the total number of runs of the circuit where a state of the form $\ket{2^i}$ was measured. This simple technique proved extremely powerful in increasing the accuracy of the algorithm.
\end{enumerate} 

We are now ready to present our experimental results.

\paragraph{Synthetic data}

For the synthetic data, we create datasets with $k$ clusters (for $k$ equal to 2 and 4) of $d$-dimensional data points (for $d$ equal to 4 and 8). This begins with generating $k$ random points that serve as the ``centroids'' of a cluster, followed by generating $n=10$ data points per centroid by adding Gaussian noise to the ``centroid'' point. The distance between the centroids was set to be a minimum of 0.3, the variance of the Gaussians was set to be 0.05 and the points were distributed within a sphere of radius 1. Therefore, there is appreciable probability of the points generated from one centroid lying closer to the centroid of another cluster. This can be seen as mimicking noise in real life data.


In Tables \ref{table_data4} and \ref{table_data8} we present, as an example, the data points and the labels assigned to them by the classical and quantum Nearest Centroid for the case of four classes in the 4- and 8-dimensional case. The first line shows how the points were sampled, namely the first quarter of points were sampled starting from the centroid labelled 0, the second quarter were sampled from the centroid labelled 1, etc. The second row shows the labels assigned by the classical Nearest Centroid algorithm. These are the labels that we will benchmark against. The third and fourth line presents the results of the quantum Nearest Centroid algorithm without and with error mitigation. We see in these examples, and in fact this is the behaviour overall, that our quantum Nearest Centroid algorithm with error mitigation provides the correct labels for 39 out of 40 4-dimensional points and for 36 out of the 40 8-dimensional points, or an accuracy of $97.5\%$ and $90\%$ respectively. 
The quantum classification is affected by the noise in the operations and can shift the assignment non-trivially based on the accuracy in the measured distance, as can be seen in cluster 1 in Table \ref{table_data8}. This will be discussed in more detail in Section E. 

\begin{figure}[H]
    \centering
    \vspace{-20pt}
    \includegraphics[width=\columnwidth]{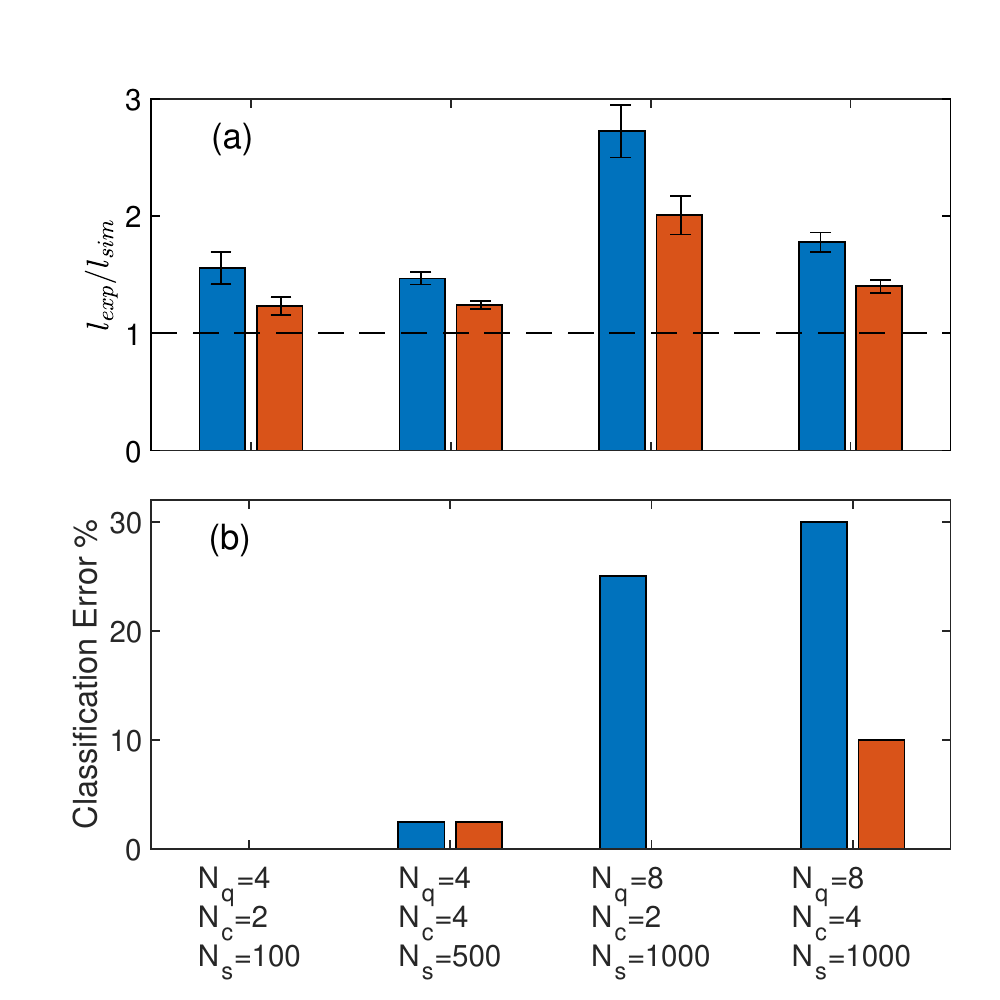}
    \caption{Synthetic data. (a) Ratio of distance calculated from the quantum computer vs the simulator, and (b) classification of synthetic data for different number of clusters ($N_c$), qubits ($N_q$) and shots ($N_s$) before (blue, left) and after error mitigation (red, right). The number of data points was 10 per cluster.  The classification error is calculated by comparing the quantum to the classical labels for each dataset.}
    \label{fig:synthetic}
\end{figure}

For the synthetic data, our goal was to see whether the quantum Nearest Centroid algorithm can assign the same labels as the classical one, even though the points are quite close to each other. Thus our classical benchmark for the synthetic data is $0\%$ classification error. The baseline which corresponds to the accuracy of just randomly guessing the labels is $1/k$ for $k$ classes.

In Fig. \ref{fig:synthetic} (b), we first show the error of the quantum Nearest Centroid algorithm averaged over a data set of 20 or 40 data points. Here, $N_c$ is the number of classes, $N_s$ is the number of shots and $N_q=d$ is the dimension and number of qubits used in the experiment. For each case, there are two values (left-blue, right-red) that correspond to the results without and with error mitigation. For 2 classes, we see that for the case of 4-dimensional data, we achieve $100\%$ percent accuracy even with as little as 100 shots per circuit and even without error-mitigation. For the case of 8-dimensional data, we also achieve $100\%$ accuracy with error mitigation and 1000 shots. We also performed experiments with 4 classes, and the accuracies were $97.5\%$ for the 4-dimensional case and $90\%$ for the 8-dimensional case with error-mitigation. The number of shots used is more for the higher dimensional case as suggested by the analysis in Section E.

While trying to estimate meaningful error bars for classification accuracy would require an impractical number of experiments (each experiment consists of running up to 160 different circuits a 1000 times each), Fig. \ref{fig:synthetic} (a) shows the average ratio between the distance estimation from experiment and simulation along with its error bars. We see that the experimentally estimated distance may actually be off by a significant factor from the theoretical distance. However, since the error bars in the distance estimation are quite small, and what matters for classification is that the ratio of the distances remains accurate, it is reasonable that we have high classification accuracy. In section E, we will show a way to reduce the error in the distance itself based on knowledge of the noise model of the quantum system.




\paragraph{The IRIS dataset}

The IRIS data set consists of three classes of flowers (Iris setosa, Iris virginica and Iris versicolor) and 50 samples from each of three classes. Each data point has four dimensions that correspond to the length and the width of the sepals and petals, in centimeters. This data set has been used extensively for benchmarking classification techniques in machine learning, in particular because classifying the set is non trivial. In fact, the data points can be clustered easily in two clusters, one of the clusters contains Iris setosa, while the other cluster contains both Iris virginica and Iris versicolor. Thus classification techniques like Nearest Centroid do not work exceptionally well without preprocessing (for example linear discriminant analysis) and hence it is a good data set to benchmark our quantum Nearest Centroid algorithm as it is not tailor-made for the method to work well. 

\begin{figure}[H]
    \centering
    \includegraphics[width=\columnwidth]{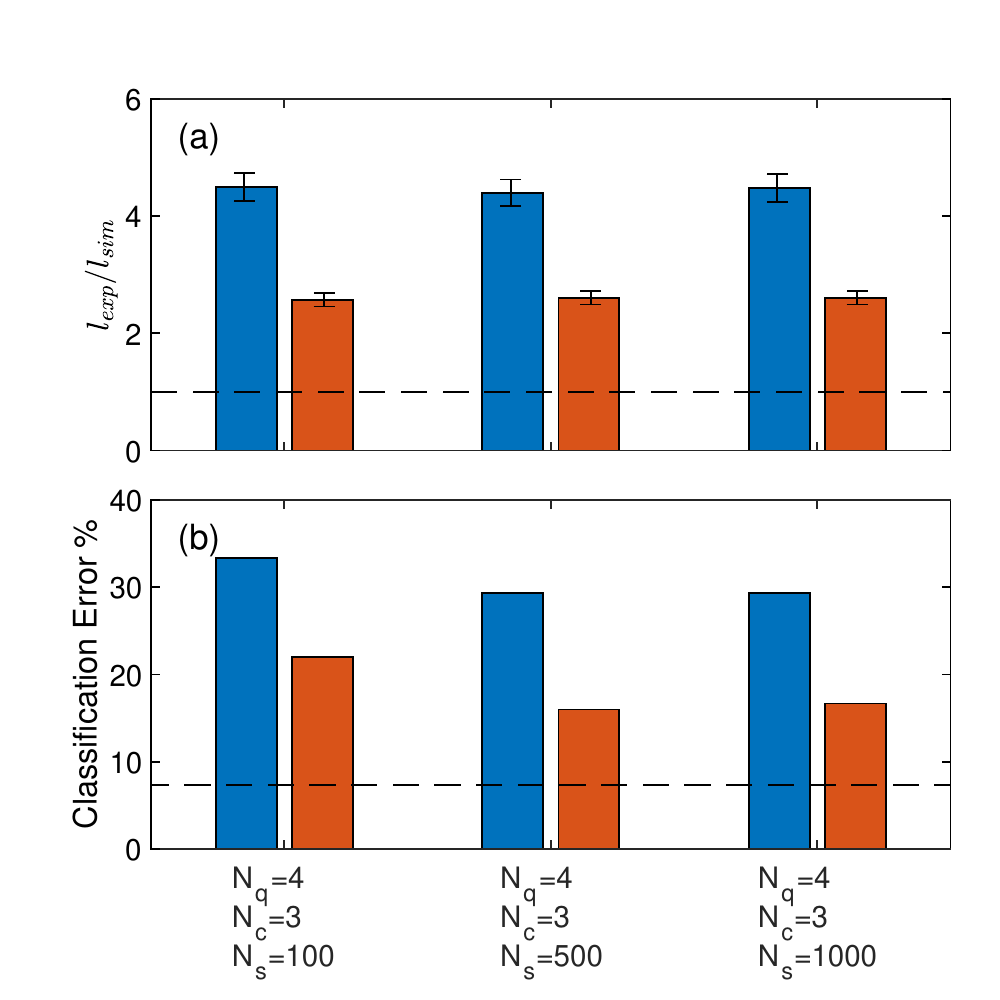}
    \caption{Iris data set. (a) Ratio of distance calculated from the quantum computer vs the simulator, and (b) classification of Iris data. There were 150 data points in total. The dashed line shows the accuracy of classical nearest centroid algorithm.}
    \label{fig:iris}
\end{figure}

\begin{figure}[H]
    \centering
    \includegraphics[width=\columnwidth]{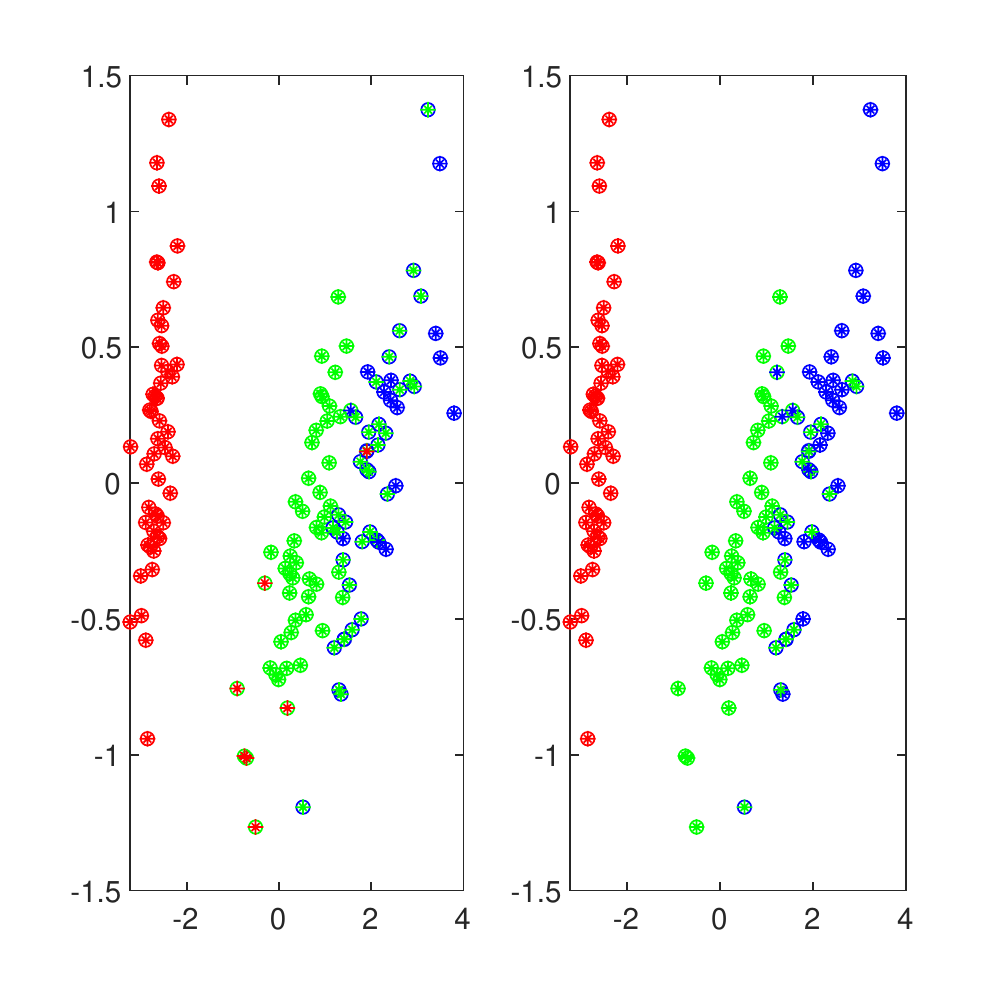}
    \caption{Iris data classification pictured after using principal component analysis to reduce the dimension from 4 to 2. The color of the boundary of the circles indicates the 3 human-assigned labels. The color of the interior indicates the class assigned by the quantum computer. (a) shows the classification before error mitigation using 500 shots, and (b) shows the classification after.}
    \label{fig:iris_data}
\end{figure}

Fig. \ref{fig:iris} shows the classification error for the IRIS data set of 150 4-dimensional data points. The classical Nearest Centroid classifies around $92.7\%$ of the points while our experiments with 500 shots and error mitigation reaches $84\%$ accuracy. Increasing the number of shots beyond 500 does not increase the accuracy because at this point the experiment is dominated by systematic noise which changes each time the system is calibrated. In the particular run, going from 500 to 1000 shots, the number of wrong classifications slightly increases, which just reflects the variability in the calibration of the system. We also provide the ratio of the experimental vs simulated distance estimation for these experiments.

Fig. \ref{fig:iris_data} compares the classification visually before and after error mitigation. Before error mitigation, many of the points that lie close to midway between centroids are mis-classified, whereas applying the mitigation moves them to the right class.


\paragraph{The MNIST dataset}

The MNIST database contains 60,000 training images and 10,000 test images of handwritten digits and it is widely used as a benchmark for classification. Each image is a $28\times 28$ image i.e. a 784-dimensional point. To work with this data set we preprocessed the images with PCA to project them into 8 dimensions. This reduces the accuracy of the algorithms, both the classical and the quantum, but it allows us to objectively benchmark the quantum algorithms and hardware on different types of data.

\begin{figure}[H]
    \centering
    \vspace{-15pt}
    \includegraphics[width=\columnwidth]{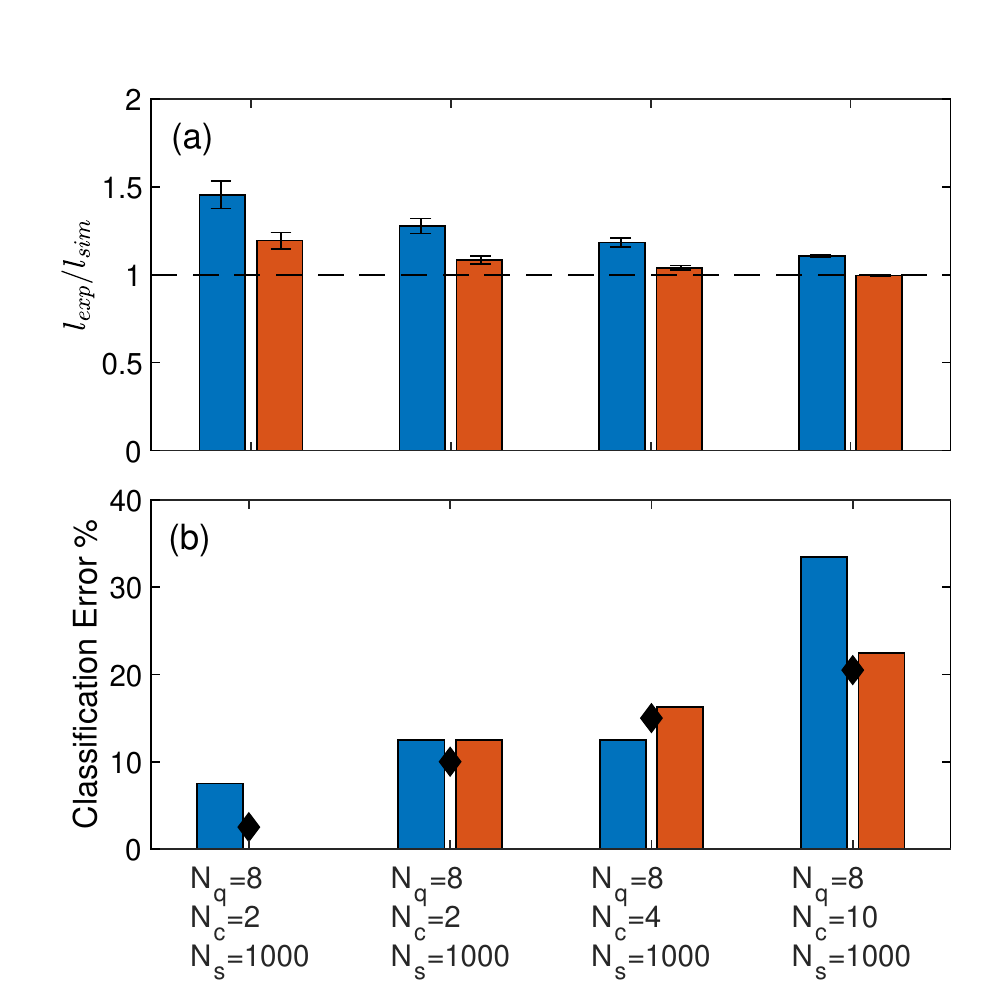}
    \caption{MNIST data set. (a) Ratio of distance calculated from experiment vs simulator, and (b) classification of different down-scaled MNIST data sets. From left to right, the data sets were (i) 0 and 1 (40 samples), (ii) 2 and 7 (40 samples), (iii) 0-3 (80 samples) and (iv) 0-9 (200 samples). The diamonds show the accuracy of classical Nearest Centroid.}
    \label{fig:mnist}
\end{figure}

\vspace{-2cm}

\begin{figure}[H]
    \centering
    \vspace{-20pt}
    \includegraphics[width=\columnwidth]{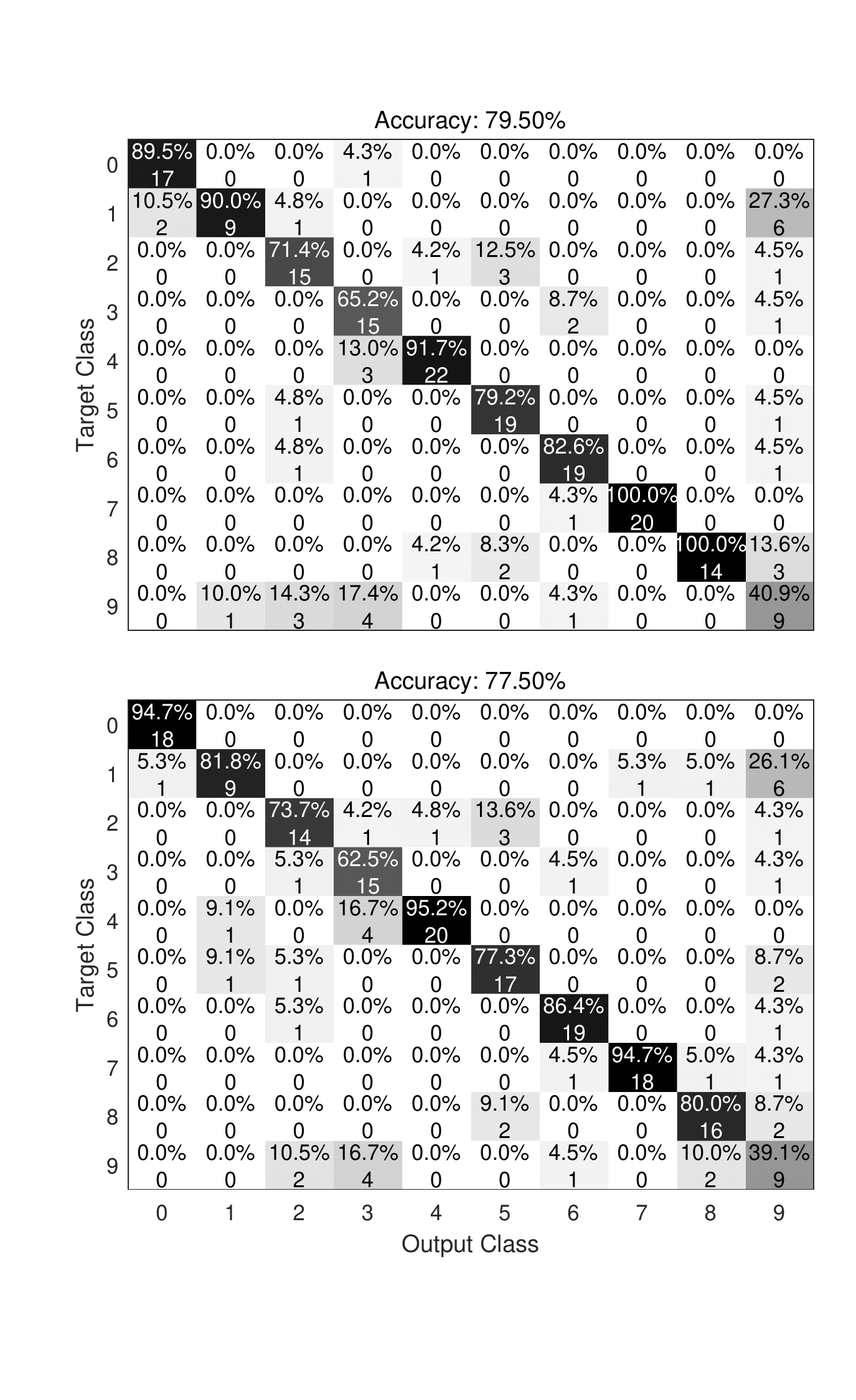}
    \vspace{-50pt}
    \caption{Confusion matrices for MNIST classification from classical (top) and quantum (bottom) computer with 10 classes and 200 points after error mitigation.}
    \label{fig:mnist_conf}
\end{figure}

For our experiments with the 8-dimensional MNIST database, we created four different data sets. The first one has 40 samples of 0 and 1 digits only. The second one has 40 samples of the 2 and 7 digits. The third one has 80 samples of four different digits 0-3. The fourth one has 200 samples of all possible digits 0-9. 

In Figure \ref{fig:mnist} we see that for the first data set the quantum algorithm with error mitigation gets accuracy $100\%$, matching well the classical accuracy of $97.5\%$. For the case of 2 vs. 7, we achieve accuracy of around $87.5\%$ similar to the classical algorithm. For the 4-class classification, we also get similar accuracy to the classical algorithm of around $83.75\%$. Most impressively, for the last data set with all 10 different digits, we practically match the classical accuracy of around $77.5\%$ with error mitigation and 1000 shots. We also provide the ratio of the experimental vs simulated distance estimation for these experiments which shows high accuracy for the distance estimation as well.

Note that this is the first time classification with 10 classes has been performed on a quantum computer with remarkably encouraging accuracy. A quantum computer of eight qubits was able to recognise eight out of ten handwritten digits on average, same as the classical algorithm, and with a time scaling which with the advent of more qubits and faster quantum clocks may become competitive with classical computers. 

In Figure \ref{fig:mnist_conf} we provide the confusion graph for the classical and quantum nearest centroid algorithm, showing how our quantum algorithm matches the accuracy of the classical one.

\subsection{Noise model and Scaling of Error Mitigation}

In this section, we discuss the different sources of error and their scaling.
One part of the effective error in the circuit can be modeled as an error that changes the state within the unary encoding subspace during two-qubit operations. This can be represented for each two qubit operation as $RBS(\theta)\to RBS(\theta(1+\Gamma r))$, where $\Gamma$ is the magnitude of the noise and $r$ is a normally distributed random number. For simplicity of the calculations we can simply say that each $RBS(\theta)$ gate performs a linear mapping for a vector $(a_1,a_2)$ in the basis $\{e_1,e_2\}$ of the form: 
\[
(a_1,a_2) \mapsto ( \cos\theta_\Gamma \cdot a_1 + \sin\theta_\Gamma \cdot a_2, -\sin\theta_\Gamma \cdot a_1+\cos\theta_\Gamma \cdot a_2)
\]
such that $\cos\theta - \cos\theta_\Gamma \approx \Gamma r$ and $\sin\theta - \sin\theta_\Gamma \approx -\Gamma r$.

It is not hard to calculate now that each layer of the circuit only adds an error of at most $\sqrt{2}\Gamma$, even though one layer can consist of up to $n/2$ $RBS$ gates. To see this, consider any layer of the distance estimation circuit, for example, the layer on time step 3 in Fig. \ref{distance}. Let the state before this layer be denoted by a unit vector $a=(a_1,a_2,\ldots,a_n)$ in the basis $\{e_1,e_2,\ldots, e_n\}$. Then, if we look at the difference of the output of the layer when applying the correct $RBS$ gates and the ones with $\Gamma$ error, we get an error vector of the form
\[
e_\Gamma = (\Gamma r_1 a_1 + \Gamma r_1 a_2, -\Gamma r_1 a_1 + \Gamma r_1 a_2,\ldots)
\]
If we look at the $\ell_2$ norm of this vector we have 
\[
\norm{e_\Gamma}^2 \leq 2\Gamma^2\norm{a}^2 = 2 \Gamma^2
\]

Since the number of layers is logarithmic in the number of qubits, this implies that the overall accuracy to leading order at the end is $(1-\Gamma)^{O(\log(n))}$. This is one of the most interesting characteristics of our quantum circuits, whose architecture and shallow depth enables accurate calculations even with noisy gates. This implies that, for example, for $1024$-dimensional data and a desired error in the distance estimation of $10^{-2}$ we need the fidelity of the $XX$ gates to be of the order of $10^{-3}$ and not $10^{-5}$, which would have been the case if the error grew with the number of gates.


Next, we also expect some level of depolarizing error in the experiment. Measurements in the computational basis can be modeled by a general output density matrix of the form
\begin{align}
    \rho_d = p \sum_{i=0}^n |a_{i\Gamma}|^2 \ket{e_i}\bra{e_i} + \frac{(1-p)}{2^n} I_{2^n},
\end{align}
where $a_{i\Gamma}$ are the amplitudes of the quantum state after the error described previously is incorporated. The depolarizing error can be mitigated by discarding the histogram states that result in states that are not $\ket{e_i}\bra{e_i}$. After this post-selection, the resulting density matrix is
\begin{align}\label{eq:depol}
    \rho = \frac{1}{\mathcal{N}}\sum_{i=1}^n \bigg(p |a_{i\Gamma}|^2 + \frac{(1-p)}{2^n} \bigg) \ket{e_i}\bra{e_i}
\end{align}

Here $p=f^m$, where $f$ is the fidelity of two-qubit gates and $m$ is the number of two-qubit gates. For our circuits, $m=4.5n-6$. The normalization factor, $\mathcal{N}= \sum_{i=1}^n \bigg(p |a_{i\Gamma}|^2 + \frac{(1-p)}{2^n}\bigg)$. Therefore, the vector overlap measured from the post-selected density matrix is
\begin{align}\label{eq:depol_ps}
    c&=\frac{p |a_{1\Gamma}|^2 + \frac{(1-p)}{2^n} }{\sum_{i=1}^n \bigg(p |a_{i\Gamma}|^2 + \frac{(1-p)}{2^n}\bigg)}\nonumber\\
    &=\frac{ |a_{1\Gamma}|^2 + \frac{(1-p)}{2^n p} }{\sum_{i=1}^n \bigg( |a_{i\Gamma}|^2 + \frac{(1-p)}{2^n p}\bigg)}\nonumber\\
    &=\frac{ |a_{1\Gamma}|^2 + \frac{(1-p)}{2^n p} }{1+\frac{(1-p)n}{2^n p}}
\end{align}

For effective error mitigation, we need $2^np \gg 1 \implies 2^n f^{4.5n-6} \gg 1$. When $n \gg 1$, this gives $ 2f^{4.5} \gg 1 \implies f\gg \frac{1}{2^{1/4.5}}=85.8\%$. Thus, we find a threshold for the fidelity over which $c\to |a_{1\Gamma}|^2$ as $n$ increases. Since the depolarizing error is much lower than this value, this implies post selection will become more effective at removing depolarizing error as the problem size increases. 

\begin{figure}
    \centering
    \includegraphics[width=\columnwidth]{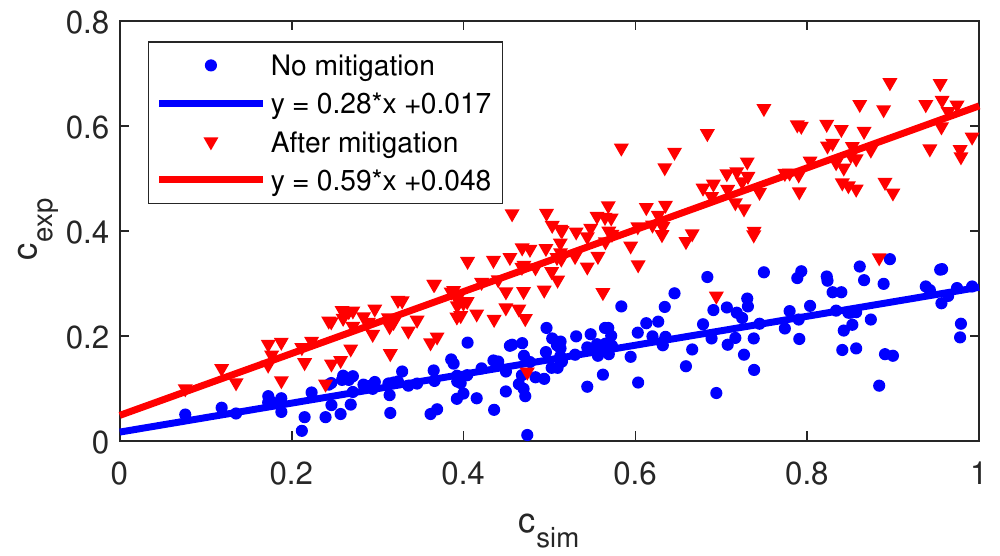}
    \caption{The experimentally estimated value of $c$ vs its value from simulation. This data corresponds to the synthetic dataset with $N_q=4$, $N_d=8$ and $N_s=1000$.}
    \label{fig:dot_prod}
\end{figure}

To test this error model, we notice that the experimentally measured value of the overlap, $c_{exp}$ should be proportional to $|a_{1\Gamma}|^2\propto c_{sim}$. We plot $c_{exp}$ vs $c_{sim}$ in Fig. \ref{fig:dot_prod} for the synthetic dataset with $N_q=8$, $N_c=4$ and $N_s=1000$ and find that the data fits well to straight lines. Using Eq. \ref{eq:depol}, we know that the slope of the line before error mitigation should be $f^m$. From this, we can estimate the value for $f$ as $95.85\%$ which is remarkably close to the expected two qubit gate fidelity of $96\%$. 

The fact that the data can be fit to a straight line can be straightforwardly used to obtain a better estimate of the distance. In this work, we have focused on classification accuracy which is robust to errors in the distance as long as they occur in the distances measured to all centroids. Nevertheless, there may be applications in which the distance (or the inner product) needs to be measured accurately, for example when we want to perform matrix vector multiplications. Note that classification is based on which centroid is nearest to the data point and thus if all distances are corrected in the same way, this will not change the classification label.

The number of samples needed to get sufficient samples from the ideal density matrix after performing the post-selection to remove depolarizing error increases exponentially with the number of qubits $n$. However, with the coming generation of ion trap quantum computers, this error will be low and thus the coefficient of the exponential increase will be extremely small. The main source of remaining error will be the first one that perturbs the state within the encoding subspace but this grows only logarithmically with $n$.

Another source of error is of statistical origin. In future experiments with much higher fidelity gates, the approximation to the distance will become better with the number of runs and scale as $1/\sqrt{N_S}$. Increasing the number of measurements is thus important for those points that are almost equidistant between one or more clusters, but is not that important for points that are clearly closer to one centroid than another. One could imagine an adaptive schedule of runs where initially a smaller number of runs is performed with the same data point and each centroid and then depending on whether the nearest centroid can be clearly inferred or not, more runs are performed with respect to the centroids that were near the point. We haven't performed such optimizations here since the number of runs remains very small. As we have said, with the advent of the next generation of quantum hardware, one can apply amplitude estimation procedures to reduce the number of samples needed. 

\section{Discussion}

We presented an implementation of an end-to-end quantum application in machine learning. We performed classification of synthetic and real data using QCWare's quantum Nearest Centroid algorithm and IonQ's 11-qubit quantum hardware. The results are extremely promising with accuracies that reach $100\%$ for many cases with synthetic data and match classical accuracies for the real data sets. In particular, we note that we managed to perform a classification between all 10 different digits of an MNIST data set down-scaled to 8-dimensions with accuracy matching the classical nearest centroid one for the same data set. To our knowledge such results have not been previously achieved on any quantum hardware and using any quantum classification algorithm.

We also argue for the scalability of our approach and its applicability for NISQ machines, based on the fact that the algorithm uses shallow quantum circuits that are provably tolerant to certain levels of noise. Further, the particular application of classification is amenable to computations with limited accuracy since it only needs a comparison of distances between different centroids but not finding all distances to high accuracy.


For our experiments, we used unary encoding for the data points, namely using one qubit per feature of the data points. While this increases the number of qubits needed to the same as in the quantum variational methods, it also allows for effective error mitigation and in particular has the advantage that the error grows only as the depth of the circuit, which is logarithmic in the number of qubits and not as the number of qubits or gates. This allows us to be optimistic for applying this algorithm to the next generation of quantum computers with a much higher number of qubits. With 99.9\% fidelity of IonQ’s newest 32 qubit system, we are confident we will match classical accuracy in higher-dimensional datasets without error correction. 

One of the useful outcomes of this work is also the development of a model of how experimental error affects algorithmic accuracy. We hope that this will spur further research on algorithmic performance on real hardware encouraging the development of practical quantum algorithms and benchmarks.

The question of whether quantum machine learning can provide real world applications is still wide open but we believe our work brings the prospect closer to reality by showing how joint development of algorithms and hardware can push the state-of-the-art of what is possible on quantum computers.

\section{Acknowledgements}
This work is a collaboration between QCWare and IonQ. We acknowledge helpful discussions with Peter McMahon.

\bibliographystyle{unsrt}
\bibliography{bibliography}

\end{document}